\RequirePackage{ifpdf}
\documentclass[letterpaper]{article}
\usepackage{jheppub}
\usepackage{xcolor}
\usepackage{amsmath}
\usepackage{epsfig}
\usepackage{caption}
\usepackage{subcaption}
\usepackage{soul}
\pdfoutput=1

\newcommand{\roughly}[1]{\mathrel{\raise.3ex\hbox{$#1$\kern-0.85em
\lower1ex\hbox{$\sim$}}}}

\def\nn{\nonumber}

\newcommand{\be}{\begin{equation}}
\newcommand{\bee}{\begin{equation}}
\newcommand{\ee}{\end{equation}}
\newcommand{\beea}{\begin{eqnarray}}
\newcommand{\eea}{\end{eqnarray}}
\newcommand{\bea}{\begin{eqnarray}}

\def\nott#1{\setbox0=\hbox{$#1$}                
   \dimen0=\wd0                                 
   \setbox1=\hbox{/} \dimen1=\wd1               
   \ifdim\dimen0>\dimen1                        
      \rlap{\hbox to \dimen0{\hfil/\hfil}}      
      #1                                        
   \else                                        
      \rlap{\hbox to \dimen1{\hfil$#1$\hfil}}   
      /                                         
   \fi}                                         %

\def\uxsl{\hbox{/\kern-.4000em$u$}}
\def\uxslsm{\hbox{\smaller/\kern-.5600em$u$}}
\def\pxpsl{\hbox{/\kern-.5000em$p$}}
\def\epssl{\hbox{/\kern-.5600em$\epsilon$}}
\def\delsl{\hbox{/\kern-.7000em$\nabla$}}
\def\lxpsl{\hbox{/\kern-.5600em$l$}}
\def\kxpsl{\hbox{/\kern-.5600em$k$}}
\def\qxpsl{\hbox{/\kern-.3900em$q$}}

\def\pref#1{(\ref{#1})}
\def\exd{{\rm d}}

\def\cG{{\cal G}}

\def\cL{{\cal L}}

\def\cR{{\cal R}}

\def\mfa{{\mathfrak a}}

\def\mfg{{\mathfrak g}}

\def\mfp{{\mathfrak p}}
\def\mfs{{\mathfrak s}}
\def\mft{{\mathfrak t}}

\def\ssE{{\scriptscriptstyle E}}

\def\EH{{\scriptscriptstyle EH}}

\def\jbar{{\bar\jmath}}

\title{Lifting Klein-Gordon/Einstein Solutions to General Nonlinear Sigma-Models: the Wormhole Example}

\author[1]{Philippe Brax,}
\author[2,3,4]{C.P.~Burgess,}
\author[5]{and F.~Quevedo}

\affiliation[1]{Institut de Physique Th\'eorique, Universit\'e Paris-Saclay,
CEA, CNRS, F-91191 Gif-sur-Yvette Cedex, France.
}

\affiliation[2]{Department of Physics \& Astronomy, McMaster University, 1280 Main Street West, Hamilton ON, Canada.
}
\affiliation[3]{Perimeter Institute for Theoretical Physics, 31 Caroline Street North, Waterloo ON, Canada.
}

\affiliation[4]{School of Theoretical Physics, Dublin Institute for Advanced Studies,
 10 Burlington Road, Dublin, 
Ireland}

\affiliation[5]{DAMTP, University of Cambridge, Wilberforce Road,  Cambridge, CB3 0WA, UK.}

\date{\today}

\abstract{We describe a simple technique for generating solutions to the classical field equations for an arbitrary nonlinear sigma-model minimally coupled to gravity. The technique promotes an {\it arbitrary} solution to the coupled Einstein/Klein-Gordon field equations for a single scalar field $\sigma$  to a solution of the  nonlinear sigma-model for $N$ scalar fields  minimally coupled to gravity. This mapping between solutions does not require there to be any target-space isometries and exists for every choice of geodesic computed using the target-space metric. In some special situations -- such as when the solution depends only on a single coordinate ({\it e.g.}~for homogeneous time-dependent or static spherically symmetric configurations) -- the general solution to the sigma-model equations can be obtained in this way. We illustrate the technique by applying it to generate Euclidean wormhole solutions for multi-field sigma models coupled to gravity starting from the simplest Giddings-Strominger wormhole, clarifying why in the wormhole case Minkowski-signature target-space geometries can arise. 
We reproduce in this way the well-known axio-dilaton string wormhole and we illustrate the power of the technique by generating simple perturbations to it, like those due to string or $\alpha'$ corrections. 
}

\begin{document}
\maketitle
\section{Motivation}
 
Although gravity is famously the weakest known force it is also the one most feared in everyday mishaps. The astronomical number of particles in even everyday macroscopic objects makes a measurable gravitational force from the extremely feeble gravitational pull of each particle separately, given that gravitational interactions of complicated sources simply add up rather than screen.

The fact that such feeble forces are detectable makes precision tests of gravity \cite{Will:2014kxa} very sensitive probes for the existence of other new fields mediating macroscopic classical forces, even if these also couple to ordinary matter at the particle level with only gravitational strength. This makes these tests in particular sensitive to the existence of any light bosons \cite{ScalarTensorTests, EPTests} whose coupling-to-mass ratio is the same for macroscopic objects as for microscopic particles. The resulting constraints are poison for the many cosmological models that would otherwise like to use light, gravitationally coupled scalars to solve problems in late-time cosmology (as has been known for many years \cite{ScaleAnomalyCC, quintessence}). 

The continued cosmological appeal of these models has led to a broad search for new types of scalar-matter interactions for which a scalar's couplings to macroscopic objects can be much smaller than -- {\it i.e.}~`screened' relative to -- their per-particle couplings \cite{Khoury:2003aq, Khoury:2003rn, Brax:2010gi, Brax:2012gr, Damour:1994zq, Olive:2007aj} (for reviews see \cite{Joyce:2014kja, Burrage:2017qrf, Brax:2021wcv}). 

A recent line of this research similarly seeks macroscopic screening due to the mutual interactions of multiple scalars \cite{Homeopathy, Brax:2022vlf, Lacombe:2023qfx}.   A practical problem when analyzing the constraints in these models is explicitly computing the scalar fields that might be expected around a macroscopic source, since this solves a coupled set of nonlinear partial differential equations in which the mutual scalar interactions play a prominent role. That is, the kinetic part of the action for a general collection of scalars
\be \label{SkinGdef}
  S_{\rm kin} = - \frac{f^2}{2} \int \exd^4x \, \sqrt{-g} \; \cG_{ab}(\phi) \, \partial_\mu \phi^a \, \partial^\mu \phi^b \,,
\ee
is characterized by a target-space metric $\cG_{ab}(\phi)$ (and overall normalization scale $f$), whose presence complicates solving the field equations found by varying $S_\EH + S_{\rm kin}$ with respect to $g_{\mu\nu}$ and $\phi^a$ (where $S_\EH$ is the usual Einstein-Hilbert action).\footnote{Our discussion ignores any scalar potential on the grounds that this is usually negligibly small in theories where the scalars are light enough to mediate macroscopic forces between test bodies. }

Progress understanding screening requires finding solutions to these equations, though explicit solutions are known only for simple special choices for $\cG_{ab}$. One such choice is the 2D $SL(2,R)$-invariant target-space metric
\be \label{SL2RMetric}
   \cG_{ab} \, \exd \phi^a \, \exd\phi^b = \frac34 \left( \frac{\exd\tau^2 + \exd \mfa^2}{\tau^2} \right) \,.
\ee
considered in \cite{Homeopathy, Brax:2022vlf}, for which the solutions seemed to rely on the existence of target-space isometries in an important way.

Our purpose here is to describe a very general mechanism for generating exact solutions to the classical field equations for multiple gravitating scalar fields that are coupled to one another through an essentially arbitrary target-space metric $\cG_{ab}(\phi)$. For spherically symmetric (or homogeneously rolling) configurations the solution-generating technique we describe gives the general solution to the field equations, and agrees with the earlier solutions of \cite{Homeopathy, Brax:2022vlf} once restricted to their specific couplings. We also illustrate our technique to rederive and generalize known axio-dilaton wormhole solutions.

\section{Mechanism}

The procedure starts with {\it any} solution, $\{\sigma(x), \mfg_{\mu\nu}(x)\}$, to the coupled Einstein/Klein-Gordon field equations for a single real scalar field,
\be\label{KGEFEs}
    \Box \sigma = 0 \qquad \hbox{and} \qquad M_p^2 \cR_{\mu\nu} + f^2 \partial_\mu \sigma \, \partial_\nu \sigma = 0 \,,
\ee
and maps it onto a general solution, $\{ \phi^a(x) , g_{\mu\nu} \}$, to the equations obtained from the action obtained by adding $S_{\rm kin}$ from \pref{SkinGdef} to the Einstein-Hilbert action. The metric and scalar fields in the sigma-model case are given explicitly by 
\be
   g_{\mu\nu} = \mfg_{\mu\nu} \qquad \hbox{and} \qquad
   \phi^a(x) = \phi^a[\sigma(x)] \,,
\ee
where the functions $\phi^a(\mfs)$ describe {\it any} geodesic of the target-space metric $\cG_{ab}$ parameterized by arc-length $\mfs$ along the curve (as measured using $\cG_{ab}$). The construction works equally well in any spacetime dimension or in Euclidean or Minkowski signature (with Euclideanization leading to a few complications, as described in \S\ref{sec:Wormholes}).

For example, specializing to 4D and using spherically symmetric configurations $\sigma = \sigma(r)$ in the equations \pref{KGEFEs} leads to the solutions \cite{ScalarTensorSoln, Burgess:1994kq}
\be \label{4DSphericallySymMetric}
   g_{\mu\nu} \, \exd x^\mu \exd x^\nu = - f(r) \, \exd t^2 + \frac{\exd r^2}{f(r)} + h(r) \, \Bigl( \exd \theta^2 + \sin^2 \theta \, \exd \varphi \Bigr)^2 
\ee
with
\be \label{SphericallySymFns}
    f(r) = \left( 1 - \frac{r_0}{r} \right)^\delta \,, \quad h(r) = r^2 \left( 1 - \frac{r_0}{r} \right)^{1-\delta} 
      \quad \hbox{and} \quad
   e^{\sqrt2\, f \sigma(r)/M_p} = \left( 1 - \frac{r_0}{r} \right)^\gamma \,,
\ee
where the parameters $\delta$ and $\gamma$ both lie in the interval $(0,1)$ because they must satisfy $\delta^2 + \gamma^2 = 1$. The corresponding sigma-model solution has precisely the same metric \pref{4DSphericallySymMetric} and \pref{SphericallySymFns} with $\phi^a = \phi^a[\sigma(r)]$ obtained by replacing $\mfs \to \sigma(r)$ in any of the target space's geodesics $\phi^a(\mfs)$. 

In the specific $SL(2,R)$-invariant case considered in \cite{Homeopathy}, where $\cG_{ab}$ is given by \pref{SL2RMetric} geodesics are given explicitly by
\be
    \tau(\sigma) = \frac{\beta}{\cosh \sigma} \qquad \hbox{and} \qquad
    \mfa(\sigma) = \mfa_0 - \beta \tanh \sigma \,.
\ee
Exact solutions to the full Einstein/sigma-model field equation are then given by using in these expressions $\sigma \to \sigma(r)$ given by \pref{SphericallySymFns}. These solutions generalize the weak-field solutions of \cite{Homeopathy} to the domain of strong gravitational fields.

The explicit and constructive nature of the solution lends itself to a broad class of applications. For instance for supersymmetric sigma-models in 4D the scalar fields are typically complex, $\{ \phi^i , \bar\phi^{\bar\imath} \}$, and with a target-space metric that is K\"ahler and so $\cG_{i \bar\jmath} = \partial_i \partial_{\bar \jmath} K$ for some K\"ahler potential $K(\phi,\bar \phi)$. Geodesics in this case satisfy the equations $\ddot \phi^i + \cG^{i\bar\jmath}\partial_k \partial_l \partial_{\bar\jmath}K \dot\phi^k \dot\phi^l = 0$. In many applications (such as to string compactifications) $K$ is given in successive approximations, but for any given accuracy of $K$ the above construction of explicit classical solutions goes through (we describe a simple example below).

The rest of this section proves the fundamental assertion underlying the above solution-generating technique. This is followed in \S\ref{sec:Wormholes} by its application to the construction of wormhole solutions for several types of sigma models coupled to gravity.

\subsection{A broad class of sigma-model solutions}
\label{sec:NewSolutions}

This section defines the system whose field equations are to be solved and describes a broad class of vacuum solutions to them.

\subsubsection{Action and field equations}
\label{ssec:ActionFE}

Consider a general sigma model containing $N$ fields, $\phi^a$, with a target space metric $\cG_{ab}(\phi)$, in terms of which target-space proper distance is given by
\be \label{GenSigModMetric}
   \exd \mfs^2 = \cG_{ab}(\phi) \, \exd \phi^a \, \exd \phi^b  \,.
\ee
We imagine the scalar self-interactions to be governed by the sigma model based on this metric, with Einstein-frame lagrangian density
\be \label{STaction}
   \cL = - \sqrt{-g} \left[ \frac{M_p^2}2 \cR + \frac{f^2}{2} \, \cG_{ab}(\phi) \, \partial_\mu \phi^a \, \partial^\mu \phi^b \right] \,,
\ee
for which the scalar field equations are
\be \label{ScalarFEGenForm}
    \partial_\mu \Bigl( \sqrt{-g} \, \cG_{ab} \,\partial^\mu \phi^b \Bigr) - \frac12\, \partial_a \cG_{bc} \, \partial_\mu \phi^b \, \partial^\mu \phi^c = \sqrt{-g} \; \cG_{ab} \Bigl[ \Box \phi^b + \Gamma^b_{cd} \, \partial_\mu \phi^c \, \partial^\mu \phi^d \Bigr] = 0 \,,
\ee
with $\Gamma^a_{bc}(\phi)$ being the Christoffel symbols built from the metric $\cG_{ab}$. The trace-reversed Einstein equations similarly are
\be \label{EinsteinGCase}
   \cR_{\mu\nu} + \frac{f^2}{M_p^2} \; \cG_{ab} \, \partial_\mu \phi^a \, \partial_\nu \phi^b = 0 \,.
\ee

\subsubsection{Geodesic solutions}

There is a very general solution to these equations obtained by specializing to $\phi^a(\sigma)$ lying along a curve in the target space along which the single variable $\sigma(x)$ varies in spacetime. When this is true the derivatives become
\be
  \partial_\mu \phi^a = \dot \phi^a \, \partial_\mu \sigma \quad \hbox{and} \quad
  \partial_\mu\partial_\nu \phi^a = \ddot \phi^a \, \partial_\mu \sigma \, \partial_\nu \sigma + \dot \phi^a \, \partial_\mu \partial_\nu \sigma \,,
\ee
where over-dots denote differentiation with respect to $\sigma$ along the curve. These imply
\be
  \sqrt{-g} \; \Box \phi^a = \partial_\mu \Bigl( \sqrt{-g} \; \partial^\mu \phi^a \Bigr) = \partial_\mu \Bigl( \sqrt{-g} \; \dot\phi^a \, \partial^\mu \sigma \Bigr) = \sqrt{-g} \Bigl( \ddot \phi^a \partial_\mu \sigma \, \partial^\mu \sigma + \dot \phi^a \, \Box \sigma \Bigr) \,,
\ee
and so the left-hand sides of the scalar field equations \pref{ScalarFEGenForm} become
\be
  \Box \phi^a + \Gamma^a_{bc} \partial_\mu \phi^b \, \partial^\mu \phi^c = \Bigl( \ddot \phi^a + \Gamma^a_{bc} \, \dot \phi^b \dot \phi^c \Bigr) \partial_\mu \sigma \,\partial^\mu \sigma + \dot \phi^a \, \Box \sigma \,.
\ee
A sufficient condition for the field equation to be satisfied therefore is to choose the curve $\phi^a(\sigma)$ to be any affinely parameterized geodesic, for which
\be \label{GeodesicCondition}
  \ddot \phi^a + \Gamma^a_{bc} \, \dot \phi^b \dot \phi^c = 0  \,,
\ee
provided we also take the spacetime-dependence of $\sigma(x)$ to satisfy $\Box \sigma = 0$.

The Einstein equations also simplify for these solutions. Parameterizing the geodesic using arc length along the curve, $\sigma = \mfs$, implies \pref{EinsteinGCase} becomes
\be \label{EinsteinGCaseMapped}
   \cR_{\mu\nu} + \frac{f^2}{M_p^2} \; \cG_{ab} \, \dot \phi^a \, \dot \phi^b \partial_\mu \sigma \, \partial_\nu \sigma = \cR_{\mu\nu} + \frac{f^2}{M_p^2} \;  \partial_\mu \sigma \, \partial_\nu \sigma = 0 \,,
\ee
where the first equality uses $\cG_{ab} \dot \phi^a \dot\phi^b = 1$ for any curve parametrised by its arc-length $\mfs$. Any target-space geodesic promotes any solution $\mfs(x)$ to the coupled Einstein/Klein-Gordon equations for a massless field into a fully nonlinear solution to the coupled sigma-model/Einstein equations.

This construction does not usually provide the general solutions to these field equations, but does so in the special case that we seek solutions only of a single coordinate -- {\it e.g.}~spherical symmetry, $\mfs = \mfs(r)$, or homogeneous time evolution, $\mfs = \mfs(t)$) -- since in this case the solution is determined by the initial values of the fields and their first derivatives in this one coordinate direction and these map perfectly onto the choices $\phi^a_0$ and $\dot \phi^a_0$ that specify a geodesic.

In the special case that the target space has symmetries then the existence of conservation laws along the geodesics makes it simpler to explicitly solve for them. For instance suppose there is a target-space Killing vector $X^a(\phi)$ satisfying
\be
     \nabla_a X_b + \nabla_b X_a = 0
\ee
where $X_a := \cG_{ab} \, X^b$. Then
\be
   \frac{\exd}{\exd \mfs} \Bigl(  X_a \dot \phi^b \Bigr) = \partial_c X_a \dot \phi^c \dot \phi^a + X_a \ddot \phi^a = \Bigl( \nabla_c X_a \Bigr) \dot \phi^c \dot \phi^a + X_a \Bigl( \ddot \phi^a + \Gamma^a_{bc} \dot\phi^b \, \dot\phi^c \Bigr) = 0 \,,
\ee
and so both $X_a \dot \phi^a$ and $\cG_{ab} \, \dot\phi^a \dot\phi^b$ are functions involving only first derivatives of $\phi^a$ that are constants along the geodesic.

\subsection{Two-field special case}
\label{ssec:TwoFieldCase}

For concreteness' sake this section explores the simplest case we explore in detail, for which there are two scalars, $\{\phi^a \} = \{ \phi, \mfa \}$. For simplicity (and with axion applications in mind) we take a target space metric for which one direction is a would-be axion, with a shift symmetry of the target space metric, and the most general metric consistent with this assumption can be written
\be \label{TwoFieldTSMetric}
   \exd \mfs^2 = \exd \phi^2 + W^2(\phi) \, \exd \mfa^2
\ee
through an appropriate choice of field variables. Stability requires $W$ does not vanish for any $\phi$. 

The gravity-scalar action in these variables then is
\be \label{TwoFieldMinkS}
   S =  - \frac{1}2 \int \exd^4x \, \sqrt{-g}\; \Bigl\{M_p^2\, \cR + f^2 \Bigl[ (\partial \phi)^2 + W^2(\phi) \, (\partial \mfa)^2 \Bigr] \Bigr\} \,,
\ee
leading to the scalar field equations 
\be \label{AxioDilatonFE}
   f^2\; \nabla_\mu \Bigl( W^2 \, \partial^\mu \mfa \Bigr)    = f^2 \Bigl[ \Box \phi - W W' \; (\partial \mfa)^2 \Bigr] = 0 \,,
\ee
and the trace-reversed Einstein equation 
\be \label{Einstein2fFE}
   \cR_{\mu\nu} + \frac{f^2}{M_p^2} \Bigl[ \partial_\mu \phi \, \partial_\nu \phi + W^2 \, \partial_\mu \mfa \, \partial_\nu \mfa \Bigr]   = 0 \,.
\ee
In particular, solutions with constant $\phi = \phi_0$ are only consistent with nonzero $(\partial \mfa)^2$ if $W'(\phi_0) = 0$.

\subsubsection{The geodesic solutions}
\label{ssec:GenSolns}

We now display the solutions to these field equations \pref{AxioDilatonFE} and \pref{Einstein2fFE} corresponding to evaluating along a target-space geodesic.

\subsubsection*{Target-space geodesics}

If we use $\phi$ as the coordinate along a curve then the target-space distance $\mfs$ between two points measured along a curve $\mfa(\phi)$ is given in terms of the target-space metric \pref{TwoFieldTSMetric} by
\be
   \mfs = \int_{\phi_0}^{\phi_1} \exd \phi \; \sqrt{\vphantom{W^A} 1 + W^2 \, ( \mfa')^2 }
\ee
where $\mfa' = \exd \mfa/\exd \phi$. Geodesics are the curves for which this is stationary against small variations of the function $\mfa(\phi)$, and so
\be
  \frac{\exd}{\exd \phi} \left[ \frac{ W^2 \mfa'}{\sqrt{1 + W^2 (\mfa')^2}} \right] = 0 \quad\hbox{which implies} \quad
  \frac{W^2 \mfa'}{\sqrt{1+W^2 (\mfa')^2}} = C
\ee
for some integration constant $C$. Choosing the convention $W > 0$ implies $\mfa(\phi)$ satisfies
\be \label{DilAxMink}
  \frac{\exd\mfa}{\exd \phi} = \frac{C}{W\sqrt{W^2 - C^2}} \quad\hbox{and so} \quad
  \frac{\exd \mfs}{\exd \phi} = \sqrt{1 + W^2 (\mfa')^2} = \frac{W}{\sqrt{W^2-C^2}} \,.
\ee
Switching to $\mfs$ as the parameter along the curve -- so $\mfa = \mfa(\mfs)$ and $\phi = \phi(\mfs)$ -- then gives
\be \label{mfaphieqs}
   \frac{\exd \mfa}{\exd \mfs} = \frac{C}{W^2} \quad \hbox{and} \quad
   \frac{\exd \phi}{\exd\mfs} = \sqrt{1 - \frac{ C^2}{W^2}} \,.
\ee
These can be integrated explicitly once $W(\phi)$ is specified.

Consider now allowing $\mfs = \mfs(x)$ to vary in space and time. Chasing through the definitions shows $\mfa(\mfs)$ inherits the derivatives
\be
   \partial_\mu \Bigl( \sqrt{-g} \; W^2 \partial^\mu \mfa \Bigr) = C \partial_\mu \Bigl( \sqrt{-g} \; \partial^\mu \mfs \Bigr)
\ee
and $\phi(\mfs)$ similarly satisfies
\be
  \Box \phi =  \frac{C^2 W'}{W^3} \, (\partial \mfs)^2 + \Box \mfs \, \sqrt{1 - \frac{C^2}{W^2}} =WW' \, (\partial \mfa)^2 +  \Box \mfs \, \sqrt{1 - \frac{C^2}{W^2}} \,.
\ee
These show that the vacuum field equations \pref{AxioDilatonFE} are automatically satisfied if $\Box \mfs = 0$. Furthermore the vacuum Einstein equation \pref{Einstein2fFE} becomes
\be \label{EinsteinKGM}
  \cR_{\mu\nu} + \left[ \left( \frac{\exd \phi}{\exd \mfs} \right)^2 + W^2 \left( \frac{\exd \mfa}{\exd \mfs} \right)^2 \right] \partial_\mu \mfs \, \partial_\nu \mfs = \cR_{\mu\nu} + \partial_\mu \mfs \, \partial_\nu \mfs = 0 \,,
\ee
where the first equality uses \pref{TwoFieldTSMetric}. As expected, the geodesic map $\{\mfa(\mfs) \,,\phi(\mfs) \}$ promotes any fully nonlinear solution $\{\mfs(x), g_{\mu\nu}(x)\}$ to the field equations for a massless Klein Gordon field coupled to Einstein gravity to a fully nonlinear solution $\{\mfa(x), \phi(x), g_{\mu\nu}(x)\}$ for the two-field sigma model coupled to gravity.

A special case of this was used in \cite{Brax:2022vlf} for the specific $SL(2,R)$ invariant choice for $W$ coming from string theory, generalizing the earlier broad class of radial solutions found for that model in \cite{Homeopathy}. But the arguments above shows that the relation is much more general because it works for more than just two fields and for arbitrary target space metrics.

\section{Applications to wormholes}
\label{sec:Wormholes}

Wormholes provide a specific and concrete example where this construction might be used. This is because explicit wormhole solutions are known for the massless Einstein/Klein-Gordon system, whose symmetries often ensure they depend only on a single coordinate. In the specific instance of wormholes the relation between solutions and target-space geodesics has been known for some time (see for example \cite{Arkani-Hamed:2007cpn, Hebecker:2018ofv}).

\subsection{Giddings-Strominger wormhole}

The simplest example starts from a Euclidean solution for a 2-form Kalb-Ramond field $B_{\mu\nu}$ coupled to gravity \cite{Giddings:1987cg} (see also \cite{Hebecker:2018ofv, Andriolo:2022rxc})
through the action
\be \label{FluxAction}
   S = -\frac12 \int \exd^4x \, \sqrt{-g} \left[ M_p^2 \cR + \frac{1}{3!f^2} \, H_{\mu\nu\lambda} H^{\mu\nu\lambda} \right]
\ee
where $H = \exd B$ is the 3-form Kalb-Ramond field strength.

The asymptotically flat Euclidean-signature field equations for this system support a wormhole solution of the form
\be \label{GSWormhole}
   \exd s^2 = \left( 1 - \frac{r_0^4}{r^4} \right)^{-1} \exd r^2 + r^2 \exd \sigma_3^2
 \quad \hbox{and} \quad
  H_{ijk} 
  = \frac{n}{2\pi^2r^3} \, \epsilon_{ijk} \,,
\ee
where $x^i$ and $\exd \sigma_3^2$ are respectively coordinates and volume element for the unit 3-sphere and $\epsilon_{ijk}$ is the volume form for the spatial slices at fixed $r$. $n$ is an integer that determines the quantization of flux
\be
   \oint_{S_3} H = n
\ee
in terms of which the field equations determine the constant $r_0$ to be
\be \label{r0vsn}
   r_0^4  = \frac{n^2}{24\pi^4f^2M_p^2} \,.
\ee
This solution with $r_0 \leq r < \infty$ describes the nucleation of the 3-sphere baby universe from a larger space, with subsequent classical baby-universe evolution starting with vanishing initial time derivatives and the initial spatial metric and 3-form field as given by the $r = r_0$ slice of the wormhole solution.

\subsubsection*{Dual scalar wormhole}

The above solution also points to the existence of a wormhole solution for the Einstein/Klein-Gordon system that can be obtained by dualizing, with dimensionless scalar $\mfs(x)$ related to 3-form field by
\be \label{HmfsDuality}
   H^{\mu\nu\lambda} = f^2\epsilon^{\mu\nu\lambda\rho} \partial_\rho \mfs\,.
\ee
Standard arguments show that using the transformation \pref{HmfsDuality} in \pref{FluxAction} leads (in Minkowski signature) to a standard scalar action
\be \label{DualActionM}
   S = -\frac12 \int \exd^4x \, \sqrt{-g} \Bigl( M_p^2 \cR +f^2 \partial_\mu \mfs \, \partial^\mu \mfs \Bigr) \,,
\ee
as appropriate for a scalar with positive kinetic energy.

There is a subtlety in this duality when applied to Euclidean-signature wormhole solutions, however \cite{Lee:1988ge, Burgess:1989da, Brown:1989df, Coleman:1989zu, Abbott:1989jw}. The subtlety arises because in Euclidean signature applying \pref{HmfsDuality} to \pref{FluxAction} leads to
\be \label{DualActionE}
   S_\ssE = \frac12 \int \exd^4x_\ssE \, \sqrt{g} \;  \Bigl( M_p^2 \cR - f^2\partial_m \mfs \, \partial^m \mfs \Bigr) \,,
\ee
rather than \pref{DualActionM}, corresponding to {\it negative} kinetic energy. Indeed, using \pref{HmfsDuality} in the solution \pref{GSWormhole} leads to a configuration that solves the field equations
\be \label{EinsteinKGE}
   \cR_{mn} - \frac{f^2}{M_p^2} \, \partial_m \mfs \, \partial_n \mfs = 0 \quad \hbox{and} \quad g^{mn} \nabla_m \nabla_n \mfs = 0 \,,
\ee
appropriate to the Euclidean action \pref{DualActionE}. 

In particular the metric remains given by \pref{GSWormhole} and the scalar is given by $f^2 \partial_m \mfs = \frac{1}{3!} \epsilon_{mnpq} H^{npq}$ and so
\be \label{GSmfsderiv}
  \partial_r \mfs  = \frac{1}{3!f^2}\, \sqrt{g_{rr}} \, \epsilon_{ijk} H^{ijk} = \frac{n}{2\pi^2f^2r^3} \left( 1 - \frac{r_0^4}{r^4} \right)^{-1/2} \,,
\ee
which implies
\bea   \label{tau}
  \mfs(r)  = \mfs_0 + \frac{n}{4\pi^2f^2 r_0^2}  \tan ^{-1}\left(\frac{\sqrt{r^4-r_0^4}}{r_0^2}\right)
  &=& \mfs_0 +\frac{n}{4\pi^2f^2 r_0^2} \cos^{-1} \left( \frac{r_0^2}{r^2} \right) \nn\\
  &=& \mfs_0 + \sqrt{\tfrac32} \left( \frac{M_p}{f} \right)  \cos^{-1} \left( \frac{r_0^2}{r^2} \right)   \,,
\eea
where the additive constant satisfies $\mfs_0 = \mfs(r_0)$ and the second line uses \pref{r0vsn} to eliminate $n$. We see that $\mfs$ asymptotes to a constant of order $M_p/f$ in the asymptotically flat large-$r$ limit and remains bounded as $r \to r_0$ with $\oint_{S_3} \epsilon_{mnpq} \partial^q \mfs \propto \oint_{S_3} H$ also finite (by construction). Eq.~\pref{GSmfsderiv} can be seen by inspection to satisfy the massless Klein-Gordon equation $\Box \mfs = 0$ for the metric of \pref{GSWormhole} because $\sqrt{-g}\, g^{rr} \partial_r \mfs$ is a constant. 

Why is it consistent to use a solution to the Klein-Gordon/Einstein equations with the wrong-sign Euclidean kinetic term as a wormhole for the scalar that has positive energy when dualized in Minkowski signature? Consistency between the Euclidean and Minkowski-signature dualities rests on careful identification of the quantum amplitude to which the wormhole contributes in both the Kalb-Ramond theory and its scalar dual (see {\it e.g.} \cite{Burgess:1990xa} for a general discussion of the relationship between WKB states and path-integral saddle points in quantum mechanics). 

The Kalb-Ramond instanton \pref{GSWormhole} corresponds to a transition where the baby universe emerges with a specified field strength $H_{ijk}$, as can be seen from the wormhole boundary conditions at $r = r_0$. For the dual scalar field the relation \pref{HmfsDuality} shows this corresponds to computing a transition to a baby universe state for which the scalar field's canonical momentum,
\be \label{ScalarBC}
  \mfp = \sqrt{-g} \, \partial_t \mfs \,,
\ee
(where $t$ is Minkowski-signature time) takes a given real value -- as opposed to being in an eigenstate of $\mfs$ itself. If we dualize in Minkowski signature and then Euclideanize we seek solutions to \pref{EinsteinKGM} rather than \pref{EinsteinKGE}, but the presence of the time derivative in the boundary condition \pref{ScalarBC} drives $\mfs$ to be imaginary in Euclidean signature when $\mfp$ is real \cite{Burgess:1989da, Brown:1989df}, leading to a saddle point for the positive-energy Euclidean action \pref{DualActionM} on which the field $\mfs$ is imaginary. This is equivalent to seeking a real solution to \pref{EinsteinKGE}, and so is a real saddle point for a wrong-sign Euclidean action \pref{DualActionE}.

\subsection{Timelike target-space geodesics}

Our goal is to map the above wormhole solution to more general sigma models using the general geodesic transformation described in \S\ref{sec:NewSolutions}. The above Euclidean considerations modify some of the details of this mapping, but it otherwise goes through as before. With the above discussion in mind we choose to explore this mapping using real solutions to a Euclidean action for which one of the scalars has a wrong-sign kinetic term (having in mind for the sigma model the same saddle-point discussion described for the GS wormhole in \cite{Lee:1988ge, Burgess:1989da, Brown:1989df, Coleman:1989zu}).

This leads to two main changes:
\begin{itemize}
\item We require one of the target-space scalars to have wrong-sign kinetic energy in Euclidean spacetime -- as the above discussion argues corresponds to preparing a baby universe in an eigenstate of canonical momentum, as would naturally occur if that particular scalar arises as the dual of a flux field like $H_{\mu\nu\lambda}$. This is ensured if the target space metric $\cG_{ab} \, \exd \phi^a \, \exd \phi^b$ has Minkowski signature rather than the standard Euclidean signature usually required by positive kinetic energy. Because of the important role played by shift symmetry when dualizing we assume the negative eigenvalue of the metric corresponds to a symmetry (axion) direction in field space.
\item Since we wish the sigma-model field equations to be automatic consequences of $\Box \mfs = 0$ and \pref{EinsteinKGE}, we demand that the geodesic solution to \pref{GeodesicCondition} be timelike, so that $\cG_{ab} \, \dot \phi^a \, \dot \phi^b = -1$. This ensures --- through \pref{EinsteinGCaseMapped} --- that the Euclidean Einstein equation \pref{EinsteinKGE} generates solutions to the Euclidean Einstein equations for the full sigma model.
\end{itemize}

\subsubsection*{Two-field special case}

As applied to the two-field special case introduced in \S\ref{ssec:TwoFieldCase} this leads to the target-space metric
\be
   \cG_{ab} \, \exd \phi^a \, \exd \phi^b = \exd \phi^2 - W^2(\phi) \, \exd \mfa^2 \,,
\ee
which is to be contrasted with \pref{TwoFieldTSMetric}. Using $\tau$ to denote proper distance along a timelike curve $\mfa(\phi)$ in this metric we have
\be \label{tauasintegral}
   \tau = \int_{\phi_0}^{\phi_1} \exd \phi \; \sqrt{\vphantom{W^A} -1 + W^2 \, ( \mfa')^2 }
\ee
where $\mfa' = \exd \mfa/\exd \phi$. 

Geodesics are solutions to
\be
  \frac{W^2 \mfa'}{\sqrt{-1+W^2 (\mfa')^2}} = C
\ee
for some real integration constant $C$ and so $\mfa(\phi)$ and $\tau(\phi)$ satisfy
\be
  \frac{\exd\mfa}{\exd \phi} = \frac{C}{W\sqrt{C^2 - W^2}} \quad\hbox{and} \quad
  \frac{\exd \tau}{\exd \phi} = \sqrt{-1 + W^2 (\mfa')^2} = \frac{W}{\sqrt{C^2-W^2}} \,.
\ee
Inverting to find $\phi=\phi(\tau)$ allows us to use $\tau$ as the parameter along the curve. This implies the functions $\mfa = \mfa(\tau)$ and $\phi = \phi(\tau)$ solve
\be \label{mfaphieqs2}
   \frac{\exd \mfa}{\exd \tau} = \frac{C}{W^2} \quad \hbox{and} \quad
   \frac{\exd \phi}{\exd\tau} = \sqrt{\frac{ C^2}{W^2} - 1} \,.
\ee

The Euclidean Einstein equation is then
\be \label{EinsteinKGM2}
  \cR_{mn} + \left[ \left( \frac{\exd \phi}{\exd \tau} \right)^2 - W^2 \left( \frac{\exd \mfa}{\exd \tau} \right)^2 \right] \partial_m \tau \, \partial_n \tau = \cR_{mn} - \partial_m \tau \, \partial_n \tau = 0 \,,
\ee
as required, where the second equality uses the timelike nature of the geodesic. Eq.~\pref{EinsteinKGM2} ensures that the metric is always given by \pref{GSWormhole} regardless of the sigma model chosen.

We must demand $0 \leq W^2 \leq C^2$ for the existence of real solutions. We also seek solutions for which $\partial_r \phi = 0$ at $r = r_0$, so that subsequent Minkowski-signature evolution starts with the initial condition $\partial_t \phi = 0$. Although this would seem to be true automatically if $\phi = \phi_0$ were constant, we have seen -- see the discussion below \pref{Einstein2fFE} -- that such solutions only exist if the constant value $\phi = \phi_0$ is one for which $W'(\phi_0)$ vanishes.\footnote{Constant-$\phi$ solutions cannot be generated using the above arguments because their starting point \pref{tauasintegral} assumes $\phi$ can be used as a parameter along the geodesic.} 

\subsubsection*{Exponential special case}

For the special case $W(\phi) = e^{-\beta \phi/2}$ the geodesic equation for $\phi(\tau)$ can be integrated quite generally and implies
\be \label{ExpTauSoln}
   e^{-\beta \phi(\tau)} = \frac{C^2 \tan ^2\left(\frac{1}{2} \beta \tau \right)}{1+\tan ^2\left(\frac{1}{2} \beta \tau \right)} = C^2 \sin^2 \left(\tfrac{1}{2} \beta \tau \right)
\ee
where the new integration constant is absorbed into $\tau_0$ appearing in the Euclideanized Klein-Gordon solution $\tau(x)$ given by \pref{tau} (and reproduced here for convenience of reference)
\be   \label{tau2}
  \tau(r) = \tau_0 + \sqrt{\tfrac32} \left( \frac{M_p}{ f} \right)  \cos^{-1} \left( \frac{r_0^2}{r^2} \right)   \,.
\ee
Using \pref{ExpTauSoln} to integrate $\mfa(\tau)$ then gives
\be \label{ExpAxSoln}
   \mfa(\tau) =  -\frac{2 \cot \left(\frac{1}{2} \beta \tau \right)}{\beta C} + c_2 \,,
\ee
where $c_2$ is a new integration constant. 

Our boundary condition is $\partial_r \phi = 0$ at $r = r_0$, but since $\partial_r \tau$ does not vanish at $r = r_0$ we choose $\tau_0$ so that $\exd\phi/\exd \tau$ vanishes at $\tau_0$ while $\mfa(\tau)$ remains bounded. This implies $\cos\left( \frac12 \beta \tau_0\right) = 0$ and so $\tau_0 = \pi/\beta$, implying \pref{ExpTauSoln} and \pref{ExpAxSoln} become
\be \label{ExpTauSoln2}
   e^{-\beta \phi(\tau)} = e^{-\beta \phi_0}  \cos^2 \left[ \sqrt{\tfrac32} \left( \frac{\beta M_p}{2f} \right)  \cos^{-1} \left( \frac{r_0^2}{r^2} \right) \right] \,,
\ee
and
\be \label{ExpAxSoln2}
   \mfa(\tau) = \mfa_0 + \frac{2 }{\beta} \, e^{\beta \phi_0/2} \tan  \left[ \sqrt{\tfrac32} \left( \frac{\beta M_p}{2 f} \right)  \cos^{-1} \left( \frac{r_0^2}{r^2} \right) \right] \,,
\ee
in agreement with the expressions given in \cite{Giddings:1989bq}. Notice that $\phi(r)$ is only bounded -- and so $W(\phi)$ is nonzero -- for all $r > r_0$ only if $\beta < \beta_c$ where
\be \label{betacrit}
  \beta_c := \frac{2\sqrt2 \, f}{\sqrt3 \, M_p} \,.
\ee

\subsection{2D K\"ahler target space}

The real power of being able to use target-space geodesics to promote the GS wormhole to a general sigma model lies in the ability it provides to generate explicit wormhole solutions in a broad class of situations. We illustrate this by exploring a broader class of 2D target spaces that in Minkowski signature correspond to 2D K\"ahler geometries. K\"ahler geometries are of interest in this context because they are naturally generated in higher-dimensional supersymmetric theories, with the K\"ahler condition related to the condition for some supersymmetries to survive compactification.

For our purposes a K\"ahler metric is a complex manifold where there exist complex coordinates $\{ z^i, \bar z^\jbar\}$ for which the metric's only nonzero components are $g_{i \jbar}(z, \bar z) = \partial_i \partial_\jbar K(z, \bar z)$ for some real K\"ahler potential $K(z, \bar z)$. For two real dimensions there is only a single complex coordinate $\{ z , \bar z\}$ and the metric is determined by one real function 
\be
   \mfg := g_{z\bar z} = \partial \bar \partial K
\ee
for $K(z,\bar z)$. For such a geometry the only nonzero Christoffel symbol is the purely holomorphic one
\be
   \Gamma := \Gamma^z_{zz} = g^{z\bar z} \partial g_{\bar z z} = \partial \ln \mfg \,,
\ee
and its complex conjugate $\overline\Gamma := \Gamma^{\bar z}_{\bar z \bar z}$. The geodesic equation for such a manifold therefore separates into a holomorphic condition
\be \label{Geo2Dholomorphic}
    \ddot z + \Gamma^z_{zz} \dot z^2 =\ddot z + \dot z^2 \partial \ln \mfg = 0 \,,
\ee
and its (antiholomorphic) complex conjugate.

Rather than continuing in these coordinates we instead change variables to the real and imaginary parts, $z = \frac12(\mft + i \mfa)$, in order to make contact with the discussion in previous sections. For the same reason we also impose a shift symmetry for $\mfa$, and so specialize to K\"ahler potentials of the form
\be
  K(z,\bar z)= K(z+\bar z) \,,
\ee
for which $\mfg = \mfg(z + \bar z) = \mfg(\mft) > 0$. In this case the target-space line element is 
\be
  \exd \mfs^2 = 2 g_{z\bar z} \exd z \, \exd \bar z = \frac{\mfg}{2} \Bigl( \exd \mft^2 + \exd \mfa^2 \Bigr)
\ee
and so the Minkowski-signature sigma-model lagrangian is given by
\be
 \cL = - \sqrt{-g} \left[\frac{M_p^2}{2} \,\cR + 2f^2 \mfg(z+\bar z) \; \partial_\mu z \partial ^\mu \bar z\right]  =  - \frac12 \sqrt{-g} \Bigl\{ M_p^2 \cR +f^2 \mfg(\mft) \left[ (\partial  \mft)^2 + (\partial \mfa)^2 \right] \Bigr\} \,. 
\ee

Contact with earlier sections is made by changing coordinates to proper distance $\phi(\mft)$, by defining $\exd \phi^2 := \exd \mft^2 \, \mfg(\mft)$ using
\be \label{phivsmft}
   \frac{\exd\phi}{\exd \mft} =\sqrt{\mfg(\mft)} \,.
\ee
because this takes the lagrangian into the form \pref{TwoFieldMinkS} with
\be \label{WphiFrommfg}
  W^2(\phi) = \mfg[\mft(\phi)] = \left( \frac{\exd\phi}{\exd\mft} \right)^2 \,.
\ee
Given a choice for $\mfg(\mft)$ one integrates \pref{phivsmft} to obtain $\phi(\mft)$ and inverts the result to get $\mft(\phi)$ and so compute $W(\phi)$ using \pref{WphiFrommfg}. The resulting geodesics can then be computed and used to generate Minkowski-signature solutions (or Euclidean-signature wormholes) by using the result in \pref{mfaphieqs} (or \pref{mfaphieqs2}) together with the appropriate Klein-Gordon/Einstein solution.

Alternatively we may express directly the solutions in terms of the fields $\mft(\tau)$ and $\mfa(\tau)$. Using \pref{WphiFrommfg} and \pref{mfaphieqs2} we get
\be\label{toftau}
\int\frac{\mfg(\mft) \exd\mft}{\sqrt{C^2-\mfg(\mft)}} = \tau+\kappa
\ee
with $\kappa$ a constant. Given $\mfg$ we can invert this equation to obtain $\mft(\tau)$. Also, since $\exd\mfa/\exd\tau = C/\mfg$ we can then find directly $\mfa(\tau)$ after solving for $\mft(\tau)$ or explicitly:
\be\label{axionoftau}
\mfa(\mft)=C\int\frac{\exd\mft}{\sqrt{C^2-\mfg(\mft)}}.
\ee

\subsubsection*{The volume modulus}

As an illustration consider
\be \label{KLogForm}
K(z+\bar z) = - \alpha^2 \ln (z + \bar z) \,,
\ee
which in practical applications often applies to a compactification's volume modulus. The special case $\alpha = \sqrt3$ is a no-scale model \cite{NoScale}. 

For this class of K\"ahler potentials we have
\be
   \mfg = K'' = \frac{\alpha^2}{(z+\bar z)^2} \,,
\ee
and so
\be
   \frac{\exd \phi}{\exd \mft} = \frac{\alpha}{\mft} \quad \hbox{which implies} \quad
   \mft(\phi) = \alpha \,  e^{\phi/\alpha} \,,
\ee
where the integration constant is used to choose the origin in $\phi$-space to satisfy $\mft(0) = \alpha$ for later convenience. Then
\be
   W(\phi) = \frac{1}{\exd\mft/\exd \phi} =  e^{-\beta\phi/2} \quad \hbox{with} \quad
   \beta = \frac{2}{\alpha} \,.
\ee
The wormhole solutions for this model are given explicitly by \pref{ExpTauSoln2} and \pref{ExpAxSoln2} above, and are nonsingular for all $r$ provided $\alpha > \alpha_c$ where (recall \pref{betacrit})
\be
  \alpha_c :=\sqrt{\tfrac32 }\left( \frac{M_p}{f} \right)  \,.
\ee

More interestingly, string-scale corrections are known in some cases to modify the K\"ahler potential for the volume modulus from \pref{KLogForm} to
\be 
   K(z,\bar z) = - \alpha^2 \ln \left[ z+\bar z + \frac{c}{(z+\bar z)^p} \right] \simeq  - \alpha^2 \ln \Bigl( z+\bar z \Bigr) - \frac{\alpha^2 c}{(z+\bar z)^{p+1}} 
\ee
for real values $c$ and $p$ (which remains a no-scale model when $\alpha = \sqrt3$ and $p = 0$). This is to be understood as the next-to-leading part of an expansion in inverse powers of $z+\bar z \gg 1$. With this choice we have
\be
  \mfg(\mft) \simeq \frac{\alpha^2}{\mft^2} \left[1 - \frac{(p+1)(p+2)c}{\mft^{p+1}} \right]
\ee
and so
\be
   \phi(\mft) \simeq \alpha \ln \mft + \frac{\alpha(p+2)c}{2 \mft^{p+1}} \quad\hbox{which inverts to} \quad
   \mft(\phi) \simeq \alpha e^{\phi/\alpha} - \frac{(p+2)c}{2\alpha^p} \, e^{-p\phi/\alpha} 
\ee
leading to
\be
  W(\phi) \simeq e^{-\phi/\alpha} \left[ 1 - \frac{p(p+2)c}{2 \alpha^{p+1}} \, e^{-(p+1)\phi/\alpha} \right] \,.
\ee

More explicitly we can find $\mft(\tau)$ and $\mfa(\tau)$ as outlined in equations \pref{toftau} and \pref{axionoftau}, with
\be
\tau+\kappa \simeq \alpha^2 \int \frac{d\mft}{\mft\sqrt{C^2\mft^2-\alpha^2}}\left(1-\frac{(p+1)(p+2)c}{\mft^{p+1}}\right)
\ee
which integrates to give $\mft(\tau)$. For the simplest case $p=0$ we can do the integrals and get
\be
 \tau+\kappa\simeq  \alpha \tan^{-1}\left(\frac{\sqrt{C^2\mft^2-\alpha^2}}{\alpha}\right)-\frac{2c}{\mft}\sqrt{\mft^2-\alpha^2} 
 \qquad (\hbox{when } p=0)\,.
\ee
For $c=0$ the second term vanishes and the first term reproduces the result of \pref{ExpTauSoln}.  Similarly:
\be
\mfa=C\int\frac{d\mft}{\sqrt{C^2-\mfg(\mft)}}\simeq C\int\frac{\mft d\mft}{\sqrt{C^2\mft^2-\alpha^2}}\left[1-\frac{\alpha^2(p+1)(p+2)c}{2C^2 
\mft^{p+3}}\right]
\ee
For $p=0$ this gives:
\be
\mfa(\mft(\tau)) \simeq \frac{\sqrt{C^2\mft^2-\alpha^2}}{C}
\left( 1-\frac{c}{\mft}\right) \qquad (\hbox{when } p=0)\,.
\ee
These simple examples illustrate the power of the general technique.

\subsubsection*{Summary}

In summary, we describe in this note a solution-generating technique that promotes classical solutions of the fully nonlinear Einstein/Klein-Gordon equations for a single scalar field into exact solutions to the classical equations for a general non-linear sigma model involving $N$ scalar fields. It returns the general solutions in situations where the fields depend only on one single parameter. 

We illustrate the method by reproducing known wormhole solutions and extending  them for more complicated target-space metrics, such as those that encode corrections to the K\"ahler potential in supersymmetric examples. Other applications include the description of homogeneous cosmological configurations when the scalar potential can be neglected, such as in a period of kinetic-energy domination or the study of spherically symmetric configurations such as arise when exploring potential screening mechanisms mediated by multiple mutually interacting scalar fields. 

More dynamical applications would be to wave propagation and to black-hole mergers or black-hole/neutron-star mergers, where solutions to the evolution for a single Klein-Gordon pair can be lifted to more general multiple-scalar models. Although these solutions need not be the most general ones in these more dynamical settings (though could be for waves that are functions of a single variable, like $x-t$), they might capture the dominant behaviour when only a single field captures the response.

Although target-space geodesics have been used in some specific instances in the literature, we hope the generality of this solution-generating technique proves useful for the many circumstances where multiple scalar fields play a role in physics beyond the Standard Model. 

\section*{Acknowledgements}
We thank Adam Solomon for helpful conversations. CB's research was partially supported by funds from the Natural Sciences and Engineering Research Council (NSERC) of Canada. Research at the Perimeter Institute is supported in part by the Government of Canada through NSERC and by the Province of Ontario through MRI. The work of FQ has been partially supported by STFC consolidated grants ST/P000681/1, ST/T000694/1.

\end{document}